\documentclass[aps]{article}
\usepackage{geometry}                
\usepackage{graphicx}

\usepackage{amssymb}
\usepackage{epstopdf}
\usepackage{amsmath}

\usepackage{amsfonts}
\usepackage{bm}
\usepackage{hyperref}

\DeclareMathAlphabet{\mathpzc}{OT1}{pzc}{m}{it}

\title{Quantum Corrections to Classical Kinetics: the Weight of Rotation}
\author{Clifford Chafin\\\ \small{Roto9 Energy, Chapel Hill, NC 27695}\thanks{cechafin@ncsu.edu}}

\begin{document}
\maketitle
\begin{abstract}
Hydrodynamics of gases in the classical domain are examined from the perspective that the gas has a well-defined wavefunction description at all times.  Specifically, the internal energy and volume exclusion of decorrelated vortex structures are included so that quantum corrections and modifications to Navier-Stokes behavior can be derived.  This leads to a small deviation in rigid body rotation for a cylindrically bound gas and the internal energy changes associated with vorticity give deviations in the Reynolds' transport theorem.  Some macroscopic observable features arising from this include variations in the specific heat, an anisotropic correction to thermal conductivity and a variation in optical scattering as a function of the declination from the axis of local vorticity.  The improvements in magneto-optical traps suggests some interesting experiments to be done in higher temperature regimes where they are not usually employed.  It is argued that the finite lifetime of observed vortices in ultracold bosonic gases is only apparent and these volume excluding structures persist in generating angular momentum and pressure in the cloud in a non-imageable form.  
\end{abstract}
\section{Introduction}
The modern understanding of heat and thermodynamics owes its solid foundations to the kinetic theory of gases \cite{Boltzmann,Lamb}.  The atomic supposition that the apparently continuum world has a finite microscopic scale of discrete atoms in the low density regime where these are free ballistic objects (for the vast majority of the time) led to a quantitative and statistical understanding of heat.  Perturbative approaches led to a microscopic understanding of hydrodynamics and continuum mechanics.  The 20th century had numerous advances and realizations in terms of the Boltzmann equations and other perturbative approaches yielding information on  correlations due to higher densities, mixtures and extreme rates of deformation.  Famous convergence problems with these series were detected and computer simulations showed logarithmic ``tails'' in the distributions \cite{Cohen, Dorfman}.  Clever resummation techniques and mode coupling models to attempt to pull out finite results that have had, arguably, some successes in agreement but still leave conceptual justification weak.  
Recently it has been argued that the diverging corrections to the Chapman-Enskog equations can be summed in their entirety to give a capillarity term and lead to the Korteweg equations \cite{Gorban} instead of  the Navier-Stokes ones.  The conditions for fluctuations to become important and limitations that cause hydro to ``break'' are also discussed and argued among proponents of quantum limits on viscosity \cite{Schaefer, Son}.  

It is not our purpose to review this extensive work but rather to reconsider it in terms of quantum dynamics and the arguable necessity for the gas to have a well defined wavefunction description at all times.  There are formal\footnote{We so often and typically stuck with the word ``formal'' in reference to such  quantum effects, meaning that we have a procedural set of rules to obtain a desired quantity rather than a justified derivation from a unifying foundation for the theory.} corrections to gas theory and the virial theorem.  Recent work on ultracold gases and the observations of ``elliptic flow'' and normal modes of these clouds in magnetic and optical traps has led to a great deal of work on gas theory applied to such systems.  However, the classical approach to gas theory in terms of ``billiard ball'' hard spheres suffers from neglecting that the delocalization time of parcel sized packets are very small compared to the lifetime of experiments.  For solids, where the particles are in continuous contact, the large mass of the system prevents delocalization on short time scales but this is a priori not true for gases.  Ehrenfest's theorem makes a kind of correspondence for wavepackets with classical trajectories but this gets less valid as the packets get wide compared to the interparticle separation.  It is not that the center of masses no longer follow the classical paths but that quantum correlations become no longer negligible.  Rapid delocalization of packets suggests that our classical kinetics is lacking in its relevance to real gases despite its successes.  Certainly it would be favorable to build a theory of gases build from a quantum construction that is ab initio.  This is not what we will accomplish here.  Our goal is, more modestly, to introduce corrections based on the following understanding of how classical gas theory must arise from wavefunction dynamics and how some aspects of vorticity should show up to low order.  

For highly isolated systems with low internal energy the many body wavefunctions delocalize and remain as such and our formal ensembles of statistical mechanics do not give any evident connection with classical gas theory.  Indeed, for those of us who believe that the system is simply a many body wavefunction at all times (hence always a ``pure state'') it is not clear why such an ensemble would be relevant and that the system tend to any equilibration at all.\footnote{It is shown in \cite{Chafin-auto} that ensembles can be relevant when one can decompose the low energy excitations of the system into a product of 3D quasiparticle excitations on top of the ground state.  These states can then form one body pseudo-eigenstates that make sense in an ensemble of such one body states that then give a dominant contribution to a many body eigenstate from these random phases and give scattering suggested by the Kubo analysis.  When far from the ground state it is not clear when this is relevant. }  The microcanonical ensemble suggests that ``thermal states'' are some states near to many body eigenstates with negligible spread in energy $\Delta E$.  Of course, any state corresponding to an interesting dynamical flow will not be in such a state.  This leads to the evident question of what induces the genesis of thermalization in quantum systems.  This has been discussed in terms of excitations into many body photon states in \cite{Chafin-auto}.  In the case of very low energy and small systems we observe vortex-lattices for Boson gases.  We know that this cannot be an eigenstate of rotation for two reasons.  Firstly, the currents at any given location are not constant.  Secondly, the vortices have an observed ``finite lifetime.''  Additionally, the angular momentum eigenstates of an interacting gas must involve vortices that align along many body directions akin to those of phonons \cite{Chafin-I, Chafin-auto} so do not generate ordered density dislocations in the one body density function.  For these states, the correlations that seem classical are therefore artifacts that are persisting from the nature of their production, generally from rotating elliptically distorted traps that are then returned to axial symmetry.  Such gases seem to not have the property of equilibration, except for carefully evaporatively ``cooled'' ones that have no angular momentum and are close to true many body eigenstates by driving out time dependent fluctuations.  

Vortex lattices seem to be a feature restricted to such low temperature gases and liquid Helium \cite{Dolfovo}.  However, if we assume that our gas is always described by a many body wavefunction then angular momentum is either in some sort of surface density wave moving around the cloud or in vortices.  These will generally not have the correlated properties that produce observable punctures in the density of the cloud but, nevertheless, do exclude some mass density from them and have the feature of being durable topological artifacts in the many body wavefunction.  In some sense, this is analogous to heat.  Heat is a correlated excitation of the gas that creates expansion but unlike heat, angular momentum conservation forces a constraint on the number of vortices so that they cannot be so easily destroyed and created.  In N-S dynamics, variations in density and pressure do allow vortex number changes and viscosity causes them to diffuse by the vorticity transport equation.  It is our interest to find the effect of buoyancy and expansion due to these vortices in otherwise classical gas behavior.  The ability to create long lasting isolated gases can also be exploited for higher temperature ones and this is one place we hope to test these results.  

For the purposes of this article, the gases that concern us are ones at high temperature with sufficient angular momentum to have an appreciable quantum vortex density and over a time scale where the ``classical correlations'' that allow an interpretation in terms of hydrodynamics to persist over the duration of observations.  This last point is especially important because this is exactly the condition we don't expect to hold for many of the current ultracold gas experiments with rotation.  The two main features we are concerned with is the volume exclusion effect induced by vorticity and the buoyancy forces on the vorticity flow in the cloud that will be shown to drive the rotational motion away from rigid body rotation.  Naturally, we expect these effects to be generally small but can appeal to the limit of strongly interacting vortex clouds at lower temperature in the regime where vortices cease to be locked into lattices for visual confirmation of some of these effects.  The case to Tkachenko modes and strain may indicate some qualitative transition to such flow on time scales where the vortex coherence persists.

The quantum vortex lattice lifetime has always been a bit problematic.  Often, by fiat, we impose a sort of two fluid model with a classical ``halo'' of gas around the condensate that then absorbs the missing angular momentum as the vortex dissipates.  The extraordinary faith in the long time validity of the Gross-Pitaevskii equation is probably one of the reasons for this despite that no serious proof of this has ever been shown \cite{Lieb, Erdos}.\footnote{Some authors have imposed near product conditions on the solutions to derive it but this is not going to yield a reasonable persistent, thus eigenstate, solution by the arguments above.}  One can argue that the many body wavefunction function simply hides this angular momentum in many body irrotational surface waves that are not long detectable in the one body density function.  It will be interesting if scattering of low energy atomic beams through the limbs of these gases eventually validates such a correlated state of motion or the classical rigid body rotation here.  This is still a problematic endeavor since these clouds typically require large expansion to image them but this reduces the density for such scattering methods.  

In the high temperature case, a classical gas requires a 3D description in terms of a parcel averaged distribution function $f(x,v,n)$.  Correlations to this function exist and are important in BBGKY type corrections to the evolution about a thermally equilibrated state \cite{Gorban}.  What is often not recognized is that, in the quantum case, the many body correlations generally forbid such a 3D picture of the gas (just as it forbids persistent vortex lattices for the low temperature case).  For such correlations to exist they generally have to be put in place by the initial data and eventually reinforced by interactions with the outside world in a fashion reminiscent of the quantum measurement problem.\footnote{But see \cite{Chafin-meas} for a possible resolution.}  For higher temperature gases that are ``classically correlated'' so that such a description is possible, we can consider this over a period of time without such external interactions to give meaning to the quantum corrections to hydrodynamics for  some period of time.  The kinds of systems where such isolation is possible are magneto-optical traps.  
Historically there was not much work done that would quantify or detect such vortex buoyancy effects because holding gases in optical traps that have interesting spatial gradient variations is a relatively new possibility.  In general, gases are held in hard walled containers and spatial variations in density are short lived or the result of powerful rotational forces in centrifuges.  

To probe the possible source of non-rigid body flow for a high temperature gas we initially look at the case of an annulus of flow of constant density with fixed angular momentum, mass and net energy.  By a variational argument we will see that the vorticity volume exclusion leads to a migration of vorticity into the interior so that there is a radial decrease from the rigid body flow profile.  After this we will develop a general modification of hydrodynamics that allows for such local deviations from the equilibrium nondissipative stress tensor.  Finally we will consider the effects of the oriented and persistent (though uncorrelated) density changes due to the hidden vortex structure in the gas and the excluded regions about it.  These will give an anisotropy in the thermal conductivity and optical scattering.  

\section{Annular Rotation}
Vortices in ultracold bosonic gases have excluded volumes about their centers of width comparable to the ``healing length'' because the wavefunction requires a node and singular vorticity at the center.  If we turn down the interaction strength, the width of this vortex should tend to the typical thermal wavelength of the gas.  Of course, as the interaction weakens, there is less reason for the cloud to form coherent vortices that are visible in the one body density profile we observe by destructive projection at the ends of these experiments.  Nevertheless, the wavefunction nature of the gas remains and angular momentum still requires such topological defects and excluded volume regions.  These excluded regions are the source of buoyancy for the vorticity.  Naturally, we would like to get some measure on the size of such an effect for a typical gas under less exotic and even real world conditions.  

Consider the case of a uniform density annulus of inner and outer radii $R_{1}$ and $R_{2}$ of gas with a given angular momentum.  There is some centrifugal rotation but if this is small along with the width of the annulus and the temperature is large enough this is a small correction.  This model can give us a platform to investigate the effect of vortex buoyancy.  To approach this problem we can treat the system as a variational problem where the net energy, mass and angular momentum of the gas is conserved.  We propose the following velocity profile for the gas\footnote{Here we coopt the differential notation ``$dA$'' and``$dB$'' to indicate small \textit{but finite} changes in the parameters.  The necessity of this will become evident when the nonanalyticity of the function about the rigid body case becomes evident.}
\begin{equation}
v=\left(\frac{dA}{r}+(B+dB)r  \right)\hat\theta
\end{equation}
whereby the irrotational component in the annulus, $dA/r$, corresponds to vortices in the mass free interior which therefore generate no compression of the gas and generates no topologically generated volume exclusion regions that generate buoyant force in the support of the annulus.  The remaining component, $\sim r$, is the rigid body part which has uniform vorticity density and is the limiting behavior we expect from classical gas theory.  We know that the minimum kinetic energy state (hence maximal thermal energy state) of a classical gas is rigid body rotation so are going to seek small corrections due to the vortex compression of the gas.  
The thermal wavelength is 
\begin{equation}
\lambda=\frac{\hbar}{\sqrt{2\pi m k T}}.
\end{equation}
We can compute the local vorticity for such a flow from
\begin{equation}
\omega=\frac{v+r v'}{2\pi r}
\end{equation}
The density of (single winding number) vortices per area is then 
\begin{equation}
n_{v}=\omega\frac{m}{\hbar}=\frac{m}{\hbar}\frac{B+dB}{\pi}.
\end{equation}
The mass constraint yeilds
\begin{equation}
M=2\pi\int \rho(r) r dr\Rightarrow \rho=\frac{M}{\pi(R_{2}^{2}-R_{1}^{2})}
\end{equation}
where we understand $\rho=m n$ to be the mass per area (due to cylindrical symmetry of the system).  
The angular momentum constraint determines the value for $B$ and a relation between $dB$ and $dA$.  
\begin{equation}
L=2\pi\int\rho(r) v(r) r^{2}dr=\frac{B}{2}\pi\rho(R_{2}^{4}-R_{1}^{4})
\end{equation}
\begin{equation}
dA=-\frac{1}{2}(R_{2}^{2}+R_{1}^{2})dB
\end{equation}
The minimum possible kinetic energy of the classical gas is the case of rigid body rotation where all the rest of the gas is at temperature $T$.  In this case the kinetic and thermal energies are given by
\begin{align}
K_{0}&=2\pi\int\frac{1}{2}\rho(r) v(r)^{2} r dr=\frac{1}{4}\pi\rho {(R_{2}^{4}-R_{1}^{4})}B^{2}\\
W_{0}&=\frac{3}{2} N k T
\end{align}
assuming the gas is monotonic.  When we include vortex volume exclusion and buoyancy the kinetic energy will have to increase.  The increase in kinetic energy for a given fraction or irrotational contribution is
\begin{equation}
d K=\frac{1}{4}\pi\rho \left((R_{2}^{2}+R_{2}^{2})^{2}\ln(R_{2}/R_{1})-(R_{2}^{4}-R_{1}^{4})\right)
dB^{2}
\end{equation}
To get the change in the thermal energy we consider the introduction of excluded regions about vortex lines to create an adiabatic compression of the gas.  The adiabaticity condition gives $Tn^{-2/3}$ as a constant so for small changes we have that $dW=N kT\frac{dn}{n}$.  The number density (particles per area) that determines the temperature is the microscopic density rather than the macroscopic average $N/Area$.  The differential change, $dn$, in the density is then given by the change in the vortex volume excluded.  (We should note that the thermal wavelength, $\lambda$, is a function of the temperature but since this is assumed to be small we can neglect this change to lowest order.)  In the case of fully rigid body rotation, the thermal energy $W_{0}$ is computed assuming the vortex volume has already been excluded.  By introducing some irrotational flow this is reduced (due to vortices moving into the interior region of the annulus) and so we expect a decrease in internal energy $dW$ since $dn$ decreases.  

The number density of vortices per area, $n_{v}=(m/\pi\hbar)B$, the area excluded per vortex, $a=\beta \lambda^{2}$, ($\beta\sim1$) gives a thermal energy change from vortex volume is then
\begin{equation}
dW=N kT \frac{d\left(\frac{N}{\pi(R_{2}^{2}-R_{1}^{2})(1-a n_{v})}\right)}{\frac{N}{\pi(R_{2}^{2}-R_{1}^{2})(1-a n_{v})}}
\approx N kT \frac{a d n_{v}}{1-a n_{v}}
\approx \frac{ N m \beta \lambda^{2} kT }{\pi\hbar}dB.
\end{equation}
Energy conservation for the system requires
\begin{align}\label{cons}
dK+dW=0
\end{align}
and this leads to an expression for $dB$ and $dA$ as function of temperature and vorticity of the flow:
\begin{align}
dB=&-\frac{\frac{ N m \beta \lambda^{2} kT }{\pi\hbar}}
{\frac{1}{4}\pi\rho \left((R_{2}^{2}+R_{2}^{2})^{2}\ln(R_{2}/R_{1})-(R_{2}^{4}-R_{1}^{4})\right)}\\
=&-\frac{2\beta \hbar}{\pi^{2} m}\frac{  (R_{2}^{2}-R_{1}^{2})}
{ \left((R_{2}^{2}+R_{2}^{2})^{2}\ln(R_{2}/R_{1})-(R_{2}^{4}-R_{1}^{4})\right)}\\
dA=&+\frac{4\beta \hbar}{\pi^{2} m}\frac{  (R_{2}^{2}-R_{1}^{2})}
{ \left((R_{2}^{2}+R_{2}^{2})\ln(R_{2}/R_{1})-(R_{2}^{2}-R_{1}^{2})\right)}.
\end{align}
The quantity $B$ plays a role of $\Omega$ for rigid body rotation.  We can use this to give a measure of the relative size of this correction.  
\begin{align}
\frac{dB}{B}\sim -\frac{ \hbar}{ m R^{2}\Omega}
\end{align}
which suggests that, in this approximation, the correction is only appreciable for very \textit{slow} rotations.  Nevertheless, we can observe that, in ultracold gas experiments, the volume exclusion of vortices is substantial so having a general theory of hydrodynamics that includes such corrections should be considered.  We regularly observe these vortices in condensates to appear to have a finite lifetime yet the radii of these clouds do not change greatly when this occurs.  This suggests that the disappearance of these do not correspond to a loss of angular momentum but a decoherence in the wavefunction that has the volume exclusion effect persist despite its absence in the form of punctured density in the one body density function.  

Note that there are an implied two roots in the above differential condition eqn.\ \ref{cons} so that we are effectively expanding about a nonanalytic point.  Using the second root to get the minimum gives a condition for small finite corrections away from rigid body rotation.  This begs some further questions.  The usual form of hydrodynamics and viscous effects leads to rigid body rotation for the fully damped limit in an axially symmetric container.  To modify such a situation means that we must alter the stress tensor term so that the vortex induced compression and buoyancy is accounted for in determining such a torque free condition on a parcels motion.

\section{Quantum Navier-Stokes}
The previous section gave us some notion of how much we might expect rigid body rotation to fail in the equilibrium limit due to pressure enhancement due to volume exclusion due to the presence of vortices.  In an external potential or pressure gradient these vortices should also have buoyancy.  In general, gas experiments had little such relevant potentials due to the lack of ability to constrain gases in manners other than hard wall containers.  The use of optical and magnetic traps at low densities have altered the possibilities here considerably.  

We are now faced with the task of introducing a scheme to incorporate both the buoyant forces on vortical elements and include the compressive effects of vortices as they are created and destroyed as well as diffuse throughout the fluid.  The vorticity transport theorem is one possible place to start but has the unfortunate side effect that the actual momentum flow of the fluid is not a local function of the vorticity so that we are likely to end up with a nonlocal equation of motion.  Navier-Stokes is really just a momentum conservation equation where the internal forces are incorporated in a local pressure and viscous stress tensor.  The usual form of the viscous shear stress tensor for simple fluids\footnote{The bulk viscosity is ignored for the usual reasons.}
\begin{align}
T_{ij}^{(s)}=\eta\left(\nabla_{(i}v_{j)}-\frac{1}{3}\delta_{ij}\nabla\cdot v\right)
\end{align}
generates rigid body rotation for rotating disks of fluid so we presume that there will need to be some quantum correction to it that gives a new vorticity dependent set point for equilibrated motion.  As we saw in the previous section, nonanalytic behavior arose about the uncorrected classical gas theory.\footnote{The history of superconductivity was held up for a long time because of a similar nonanalytic behavior that invalidated the usual perturbative methods.}   Perturbative approaches generally assume analyticity but by not requiring such an assumption we have a larger range of possible ansatz forms from which to select such a correction.

The dimensions of $T$ and $P$ are force/area.  The buoyant force on a parcel (in force per volume) is
\begin{align}
f=-\rho\cdot(1-\beta\lambda^{2}n_{v})\cdot g-\nabla P
\end{align}
where we remember that $\rho$ is the ``microscopic'' density between the vorticity excluded density regions.  We seek to find a plausible equilibrium momentum shear as a function of $f$.  The direction of this shear should be perpendicular to the local direction of the vorticity and the gradient of this and to the force $f$ itself.  Converting the force $f$ to a stress requires a further length scale.  The mean vortex separation, $n_{v}^{-1/2}$, and the vortex size, $\lambda$, are the only such scales available to us.  The former one gives a measure of the density of vortices so is the natural choice.  Using these we can define a correction stress term
\begin{align}
t_{ij}=\zeta|f n_{v}^{-1/2}| u_{(ij)}\\
u=(\hat\omega\times\hat f)\otimes\hat f
\end{align} 
where $\zeta\sim1$ is a dimensionless constant.
The resulting stress tensor is then
\begin{align}
T_{QNS}=T_{NS}+t
\end{align}

The pressure field also needs correcting in the manner of the previous section.  
\begin{align}
P=n_{m} kT
\end{align}
where $n_{m}=\rho(1-\beta\lambda^{2}n_{v})/m$ is the microscopic particle density.  
The final quantum corrected N-S equations are then
\begin{align}\label{QNS}
\rho\frac{D}{Dt}v=\rho(\partial_{t}v+v\cdot\nabla v)=-\nabla P+\eta\nabla\cdot T^{(QNS)}
\end{align}

This is evidently a momentum conserving hydrodynamic equation however we have to consider energy conservation as well.  In usual hydrodynamics we apply the Reynolds' transport theorem to 
\begin{align}\label{energy}
\frac{D}{Dt}\int\rho(e+\frac{1}{2}v\cdot v)
\end{align}
and contract N-S with $v^{i}$ to extract the mechanical part.  Here $e$ is the internal energy per mass.  The difference gives us a measure of the internal (thermal) energy of the system.  Here we have an additional complication.  The volume exclusion of changes in vorticity occur on a microscopic scale and so are ``internal.''  To get the total energy we need to consider the internal flow associated with vorticity creation, the work this does agains the background pressure $P(\rho_{m},T)$ and the quantum curvature energy that holds the amplitude out from the vortex core.  Once these are accounted for along with the macroscopic mechanical (kinetic) energy, the deficit is the thermal energy that determines $T$ (hence $P$).  A natural length element defined by the vortex structure is $d=n_{v}^{-1/2}$ so the volume element is $n_{v}/d=n_{v}^{3/2}$ and the relative volume fraction of the vortices is $\beta\lambda^{2}/d^{2}=\beta\lambda^{2}n_{v}$.  The internal flow can be modeled by thinking of vortices as of fixed area $\beta\lambda^{2}$ and growing length along $\pm\hat\omega$ (with no preferred direction).  The internal kinetic energy density is then only present when vorticity is changing.  The velocity of the increase then defined by $\dot{d}$ so that 
\begin{align}
K_{int}=\alpha_{K}m(\dot d)^{2}n_{v}^{3/2}=\alpha_{K}m \frac{\dot{n_{v}}^{2}}{n_{v}^{3/2}}=\alpha_{K}\sqrt{\frac{m^{3}}{\hbar}}\frac{\dot\omega^{2}}{\omega^{3/2}}
\end{align}
The quantum curvature energy density is a function of the total vorticity present so is of the form
\begin{align}
U_{Q}=\alpha_{Q}\frac{\hbar^{2}}{2m\lambda^{2}}\beta\lambda^{2}n_{v}=\alpha_{Q}\frac{\beta\hbar^{2}}{2m}n_{v}=\alpha_{Q}\beta\hbar\omega
\end{align}
where $\alpha_{K}$ and $\alpha_{Q}$ are constants of  $\mathcal{O}(1)$.  
The density work done on the gas due to volume exclusion is given by the pressure times the rate of change of the excluded volume fraction
\begin{align}
\partial_{t} U_{v}= P\partial_{t}(\beta\lambda^{2}n_{v})=\frac{\beta m\lambda^{2}}{\hbar}P\dot\omega=\frac{\beta \hbar}{2\pi kT}P\dot\omega
\end{align}

To complete the closure of the equations of motion we need to determine the evolution of the internal energy, through the temperature, of the system.  The total boundary of the system has internal and external parts.\footnote{The picture of internal surfaces here is certainly idealized.  In reality, we consider that this vorticity is decorrelated so that there is no well-defined 2D classical surface for this gas.  The working assumption is that assuming such a correlated behavior gives equivalent results.  Any mention of the ``internal'' surface vanishes from the final results.}  The quantum energy is a surface contribution that is ostensibly kinetic but is analogous to a stationary quantum curvature not a dynamic high temperature transfer of energy we that diffuses freely as in classical thermodynamics.  The updated advected energy is
\begin{align}\label{energy}
\frac{D}{Dt}\int_{V}\left(\rho_{m} e_{th}+ \frac{1}{2}\rho_{m} u^{i} u_{i}+U_{Q}\right)=&\int_{V}\rho_{m} F^{i}u_{i}+\int_{\partial V}\left(\nu^{i}T_{ij}u^{j}-\nu^{i}q_{i}  \right)+\int_{V}\rho_{m} F_{(int)}^{i}u_{i}
\end{align}
where $u$ is the full velocity field including microscopic flows.  
The surface normals are given by $\nu^{i}$.  
The volumes have been selected so that the excluded core regions give internal surfaces given by $\partial V_{int}$.  This ensures that there can be no thermal flux term $q_{i}^{(int)}$ through these surfaces.  
We usually now perform integration by parts to make this a pure volume integral and invoke the condition that the arguments must be equal.  We want a more coarse grained approach.  Specifically, we would like to consider parcels that are small but large enough that the vorticity excluded regions are small and generally quite numerous in parcels.  The energy propagation expression is
\begin{align}\label{energy}
\frac{D}{Dt}\left(\rho e_{th}+ \frac{1}{2}\rho (v^{i}+v_{(m)}^{i}) (v_{i}+v_{(m),i})+U_{Q}\right)=\rho F^{i}v_{i}-\partial_{i}(P v_{i})\\+\partial_{i}(T^{(s)}_{ij}v_{j})-\partial_{i}q_{i}
+\rho_{m} F_{(int)}^{i}v_{(m),i}
\end{align}
This can be simplified by breaking up the terms into the small and large scale contributions
\begin{align}
\frac{D}{Dt}\left(\rho e_{th}+ \frac{1}{2}\rho v^{i}v_{i}+K_{int}+U_{Q}\right)=\rho F^{i}v_{i}-\partial_{i}(P v_{i})+\partial_{i}(T^{(s)}_{ij}v_{j})-\partial_{i}q_{i}
+\partial_{t} U_{v}\\
=\rho F^{i}v_{i}-v_{i}\partial_{i}P -P(\nabla\cdot v)+v_{j}\partial_{i}T^{(s)}_{ij}+(\partial_{i}v_{j})T^{(s)}_{ij}-\partial_{i}q_{i}
+\partial_{t} U_{v}
\end{align}
We can cancel the mechanical contribution by subtrating off the mechanical contribution from the momentum eqn.\ \ref{QNS}.  
\begin{align}
\frac{D}{Dt}\left(\rho e_{th}+K_{int}+U_{Q}\right)=\rho F^{i}v_{i}-P(\nabla\cdot v)+(\partial_{i}v_{j})T^{(s)}_{ij}-\partial_{i}q_{i}
+\partial_{t} U_{v}\\
\rho\frac{D}{Dt} e_{th}= -\rho\frac{D}{Dt}\left(K_{int}+U_{Q}\right)       +\rho F^{i}v_{i}-P(\nabla\cdot v)+(\partial_{i}v_{j})T^{(s)}_{ij}-\partial_{i}q_{i}
+\partial_{t} U_{v}
\end{align}
The nature of the averaging now allows us to use the macroscopic density $\rho$ instead of the microscopic $\rho_{m}$.  The velocity has now been decomposed into $u=v+v_{(m)}$ to give the velocity of bulk motion and the velocity of flow given by the increase in small scale vorticity (but not including the thermal motion which is described by $e_{th}$).  Because the directions of vortex growth are random there are no net cross terms of $v \cdot v_{(m)}$.  This establishes the evolution equation for the thermal energy density $e=\frac{3}{2}n_{m}kT$.  The heat flux term $\partial_{i}q_{i}$ is generally treated by introducing Fick's law $q_{i}=-\kappa {\partial_{i} T}$ (however see below for vorticity implied anisotropic corrections).  

There is one problem with this set of equations of motion.  Two of these terms are explicitly a function of $\dot\omega$ and $\ddot\omega$ (hence $\dot v$ and $\ddot v$).  Including this in a dynamical equation is problematic.  An analogous problem arises in the computation of the radiation reaction problem and solving the Abraham-Lorentz-Dirac equations.  The equations of motion with the lowest order corrections are
\begin{align}
\frac{d p^{\mu}}{d\tau}=F^{\mu}_{ext}+\frac{\mu_{0}e^{2}}{6\pi m c}\left(\frac{d^{2} p^{\mu}}{d\tau^{2}}-\frac{p^{\mu}}{m^{2}c^{2}}\frac{d p^{\mu}}{d\tau}\frac{d p_{\mu}}{d\tau} \right).
\end{align}
The solution to this can be found by realizing that all charged objects in electromagnetism should be bound objects of finite size.  The higher order effects we are then observing are artifacts of time delay of the self field crossing the body.  In fact, in this perspective the $m_{rest}$ mass should really be replaced by $m_{bare}$, the nonelectromagnetic mass, and the lowest order effect of the acceleration fields then give a backwards force that generate a $m_{em}a^{\mu}$ force that moves the energy associated with the velocity field of the object but in a completely local way by only acting on the localized bare mass and allowing the requisite field cancellations to alter the velocity field energy causally.  If we reinterpret the $\ddot p^{\mu}$ term as the result of a force from the gradient of the external field across the body (due to its nonlocal features) the term can be replaced by a force proportional to $p^{\mu}\nabla_{\mu}F_{ext}^{\nu}$ so that 1.\ the degrees of freedom of the system are unaltered and 2.\ the identity $v_{\mu}v^{\mu}=-c^{2}$ is preserved in a locally averaged fashion.  Higher order corrections involve the finite size parameters of the body explicitly.  

The resolution of this problem for our QNS equations are to solve the equations without the small $\dot\omega$ and $\ddot\omega$ terms and iterate the results for $\dot\omega$ and re-solve them.  This gives a closure scheme for the variables $\rho$, $v$, and $T$.

\section{Macroscopic Probes}
The above treatment was based on the conception that a gas is a many body wavefunction at all times and is sufficiently delocalized so that the packet based approach to ballistic motion is invalid.  Furthermore, the scale of the gas and its interactions with external matter are strong enough to maintain the kinds of classical correlations that allow a classical description of it in terms of macroscopic variables.  There are some dramatic consequences of this picture.  Firstly, the collisions in a classical gas are constant and should be generating radiative losses continuously.  This is not the case in a classical gas where, presumably, on suitably small parcels that are many interparticle separations wide, the wavefunction is close to an eigenstate.  Here time dependent fluctuations are small and the kinetic energy is primarily in near standing waves rather than nonzero currents.  The absence of observations of such radiation from gas samples of small optical depth is evidence for this picture.  Let us now seek some non null observations to justify it.  

The presence of an unseen and nonclassical set of vortical structures of excluded amplitude imposes an anisotropy on a gas with vorticity even if it is stationary.  Let us examine possible effects on thermal conductivity, specific heat and optical scattering.  Firstly, the volume exclusion effect alters the effective volume in a fashion analogous to the van der Waals correction to gas thermodynamics and specific heats.  This correction is elementary.  Thermal conductivity and viscosity in gases rely on the transport of  energy and momentum in a diffusive fashion controlled by the particle velocities and mean free paths.  These calculations are generally from a classical ballistic model.  It would be desirable to generate this, and the equations of state from an analysis of a many body wavefunction directly.  Some discussion of this is given in \cite{Chafin-th}.  For now let us consider a more heuristic approach.  

In quantum optics a convenient maxim is ``a photon only interferes with itself.''  This picture presumes a separable system so is not completely general but does give us enough intuition to resolve some paradoxes regarding classical versus quantum optics.  An analogous statement regarding vortices in many body wavefunctions would be ``a particle is only scattered by its own vortices.''  This again presumes a product function structure of the wavefunction but it gives us some sort of bound on how much the vortical structure in a stationary gas flow impedes thermal energy.  The 1D nature of vortices means that the scattering parallel to them should be negligible while that perpendicular to them will be obstructed by a set of columnar obstructions of 1D cross section $\sigma\sim\lambda(T)$.  The vorticity density determines the density of such structures, $n_{v}$.  As long as this is a small correction to the classical process, we can estimate the correction to thermal conductivity by considering the corrections to the mean free path.  The kinetic theory result for thermal conductivity of a monotonic gas is
\begin{align}
\kappa=\frac{1}{3}n v c l=\frac{1}{2} n v l
\end{align}
where $v$ is the thermal rms velocity and 
\begin{align}
l=(n\sigma)^{-1}
\end{align}
is the mean free path and $\sigma=4 a_{p}$ with $a_{p}$ the area of the colliding particles.  The typical trajectory of a particle will be a sphere of radius $l$.  The probability that a vortex line falls in such a sphere is $p=\pi l^{2 }n_{v}$.  The scattering then possesses another scattering possibility with area $a_{l}\approx \xi \sqrt{a_{p}} l$ (where $\xi$ is a shape factor $\sim1$).  We can then modify the mean free path for transverse motions by 
\begin{align}
l_{\perp}=(n\sigma + (n_{v})(\xi \sqrt{a_{p}} l))^{-1}.
\end{align}
We define $l_{||}$ to be the uncorrected mean free path.  The resulting thermal conductivities parallel and perpendicular to $\omega$ are
\begin{align}
\kappa_{||}&=\frac{1}{2} n v l_{||}\\
\kappa_{\perp}&=\frac{1}{2} n v l_{\perp}. 
\end{align}
We can compute the density of vortices per particle by $n_{v}=(m/\hbar)\omega\approx10^{10}\omega$.  Thus for rotating fluids with periods $\tau\sim1$s we have vortex separations $d\sim10^{-3}$m.  Since gases at atmospheric temperature and pressure have mean free paths $l\sim10^{-6}$m these will be small correction unless the gases are dilute or spinning very rapidly.  

In the case of electromagnetic scattering, there are famous results for optical scattering due to density fluctuations in liquids \cite{LL-fluid}.  If the vortex line structure suffers little lateral variation then we can assume light traveling parallel to $\omega$ will be unaltered by the presence of vortices.  For transverse light, the model we suggest is that of a dielectric with column-like cavities arranged in random patterns with an averaged density matching $n_{v}(x)$.  The frequencies most scattered will be those of wavelength comparable to $\lambda(T)$.  At room temperature, this will correspond to X-rays.  For realistic experiments our gas samples will then be optically very thin.  This lets us ignore Anderson localization effects and consider only the net number and density of vorticity regions encountered.  The power loss from scattering in a wavefunction depleted region will be $\sim \sqrt{\epsilon}-1$ for the permittivity of the gas.  Assuming $\lambda_{em}\approx\lambda(T)$, the loss for a sample of thickness $L$ should be $\Delta P_{trans}^{(\perp)}\sim (\sqrt{\epsilon}-1)n_{v}\lambda L P_{inc}$.

\section{Conclusions}
We have investigated some implications of the notion that a gas is a many body wavefunction at all times with delocalization that persists on the scale of many interparticle separations.  Specifically, the implications that angular momentum produces some microscopic volume exclusion due to many body vortex defects in the gas have effects on both the pressure of the gas and introduce anisotropy in it.  For an annular rotating flow, small irrotational corrections produce a slowdown at the outer edge of the ring and give a deviation from rigid body rotation for the equilibrated state.  This result follows from variational considerations where mass, energy and angular momentum are constrained to be constant.  Interestingly, nonanalytic behavior about the classical case exists in this analysis.  A modified Navier-Stokes set of equations are suggested based on an ansatz for the buoyancy introduced by vortices and the pressure increase due to microscopic volume exclusion of vortices.  Some complications related to bounds on the dynamical degrees of freedom arise but these can be solved by similar constructions that resolve the analogous problem and runaway solutions for the radiation reaction of charged particles.  

Macroscopic implications of this fundamentally quantum constructions range from the absence of radiation we would expect from a classical gas picture generated from collisions to $\lambda$-scale preferential radiation scattering and an anisotropic thermal conductivity (and possibly viscosity).  These features are not a feature of nonequilibrium dynamics but are merely due to the presence of vorticity so apply to equilibrated rotating gases.  Such features seem within the realm of detection with strongly driven systems so provide avenues to validate or invalidate these arguments in a reasonable time frame.  
Much of this work has a heuristic character but is meant to be evocative and illuminate the need for a fully quantum mechanical theory of gases that allows for the emergence of hydrodynamic behavior as a direct consequence of the many body Schr\"{o}dinger equation.  

This work suggests many questions for further investigation not the least of which is the generality and microscopic origin of the quantum stress correction $t_{ij}$.  Given our semi-classical picture one can ask what happens when the wavelength of heat is comparable to $\lambda$ and such a flux passes near the vortex?  Higher energy waves could pass more closely to the core so are potentially scattered more steeply.  Lower energies have the potential to be scattered with a greater cross section.  This suggests a kind of ``heat prisming'' effect that could affect thermal transport and might cause some flows to generate thermal inhomogeneities even when there is no differential heating.

%
%
%
%
%
%
%
%
%
%

\end{document}